# TIME CRITICAL MULTITASKING FOR MULTICORE MICROCONTROLLER USING XMOS® KIT

Prerna Saini[1,] Ankit Bansal[2] and Abhishek Sharma[3]

[1,3]Department of Electronics and Communication Engineering, LNM Institute of Information Technology, Jaipur, India

[2]Department of Electronics and Communication Engineering, GLA University, Mathura, India

## ABSTRACT

*This paper presents the research work on multicore microcontrollers using parallel, and time critical programming for the embedded systems. Due to the high complexity and limitations, it is very hard to work on the application development phase on such architectures. The experimental results mentioned in the paper are based on xCORE multicore microcontroller form XMOS®. The paper also imitates multi-tasking and parallel programming for the same platform. The tasks assigned to multiple cores are executed simultaneously, which saves the time and energy. The relative study for multicore processor and multicore controller concludes that micro architecture based controller having multiple cores illustrates better performance in time critical multi-tasking environment. The research work mentioned here not only illustrates the functionality of multicore microcontroller, but also express the novel technique of programming, profiling and optimization on such platforms in real time environments.*

## 1. INTRODUCTION

In the present era of technology, computational power [1] plays an important role. The multicore microprocessor devices [2] are already available in CISC architecture which used to perform non real time computing. Recently there has been a huge demand of high computing speed in time critical system, mostly in real time embedded device. Technology is growing exponentially every day with the demand of more power and processing handling capabilities. The basic need of a multicore system is the distributed and parallel computing [3]. Time consumption is the drawback of single core processors, so multicore [4] technology is used to achieve efficiency through parallel processing. Parallel processing [5] is the simultaneous use of more than one CPU to execute a program or multiple computational threads. The main goal of parallel processing is a high performance [6] computing, which speedup the execution time of the program. Parallel [7] processing makes programs run faster because there are more engines (CPUs or cores) workingon it. It increases the efficiency, safe execution time, take less energy and retain the time. Multicore has two or more CPUs while the single core has only one core inside it as illustrated in figure1. To enhance the performance [8] of single core processor, it is mandatory to increase the frequency as CPU load increases. It causes heat losses and leakage current so rather than increase the clock frequency of single core, manufacture switched to multicore to avoid the power [9] consumption problem and to increase speed and efficiency.

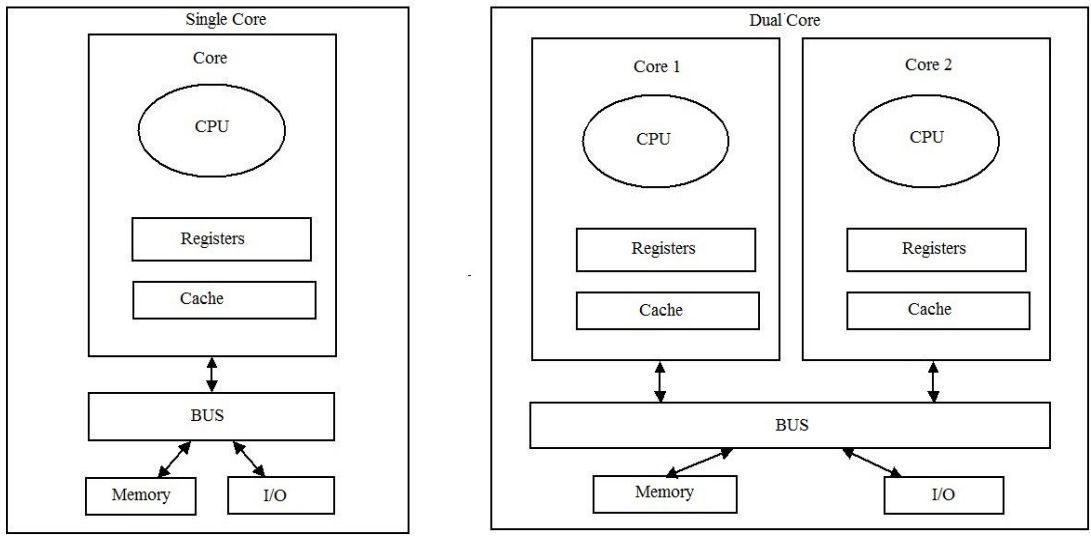

Figure 1. Block Diagram of Single-core and Multi-core Processor

As the number of the taskis rapidly increasingthe user wants to perform more than one task at a time, but a computer with a single-core performs one operation at a time [10]. Although with software threads, some amount of parallelization is possible, but it does not give satisfactory results. In multicore scenario, it is possible to perform operations at comparatively high speed [11] to perform paralleled task and save time. Table 1 represents the difference between single core and multicore and shows that multicore has more advantage over single core with respect to processing speed, power and operation handling ability etc.

Table 1. Comparison between single core and multi core system

| Parameter | Single-Core | Multi-Core |
|---|---|---|
| No of cores | One primary core | Two or more separate core |
| Processing | Sequential | Parallel |
| SMT | Not Possible | Possible |
| Power | Low | High |
| Speed | Slow | Fast |
| Efficiency | Low | High |
| Operation | One task at a time | Multitasking |

The multi-processing system has two types, namely homogeneous and heterogeneous. If all the cores are identical and have the same features like message passing system, cache, threading, share memory and resources then it is called homogeneous multi-core processors. In heterogeneous [12] systems all the cores have different features; it can vary clock cycles according to system requirements to achieve low power or ultra-low power mode.Flynn's taxonomy is a specific classification of parallel computer architectures that are based on the number of concurrent instruction (single or multiple) and data streams (single or multiple) available in the architecture [13]. Figure2 shows the evolution of multicore era thatevery single processor replaced by a multicore processor to get high performance [14]. At the present time, there is vast use of multicore [15] system. In future, no filed will be untouched with multicore system.The first dual core microprocessor was Power-4 [16], [17] which was designed by the IBM® in 2001.

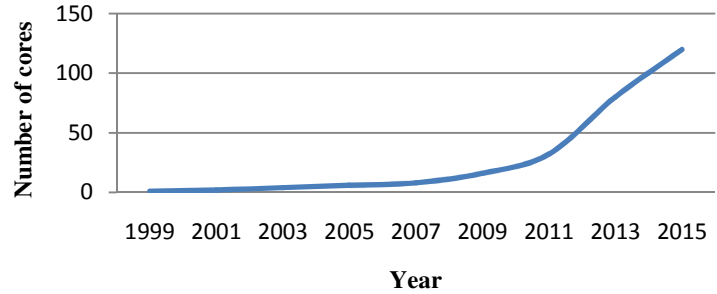

Figure 2. Evaluation Era of Multicore system

Amdahl's law [18] is used to find out the speedup of a multicore system. Speedup is how much time taken by a program to execute through the single core divide by the time taken by a program when n number of processors execute in parallel manner as illustrated in equation 1 and 2 which shows that as the number of cores in a processor is increased, the speed of a system is also enhanced, but it is impossible to fully parallelize the program. Suppose a program takes 10 hours using single core and a particular portion of that program which cannot be parallelized take at least 1 hour, then by increasing the number of cores, execution time and speedup of the system cannot be changed. By increasing the number of cores somehow speed up is increased, but it depends upon how much a program can be parallelized.

$$T(n) = T(1)\left(B + \frac{1}{n}(1-B)\right) \qquad \dots (1)$$

$$S(n) = \frac{T(1)}{T(n)} = \frac{T(1)}{T(1)\left(B + \frac{1}{n}(1-B)\right)} = \frac{1}{B + \frac{1}{n}(1-B)} \qquad \dots (2)$$

In parallel processing [19], every core executes the multiple set of instructions as an individual processing unit. The CPU istreated as a single unitso the end user can divide the whole task into subsection and send to various cores. Due to parallel processing all the core work simultaneously so it enhances processing speed and save the time. Important of parallelism is increased because complex problems can be split into smaller programs that can be executed at the same time to reduce execution time. Basically parallelism is subdivided in two broad areas i.e. Instruction level parallelism (ILP) and threads level parallelism. ILP tells, how many simultaneous instructions can be executed, and thread level parallelism (TLP), is about how many simultaneous threads can be executed. In a multicore system Parallelization is possible because it spilt threads and assigns it to each core and all works simultaneously. Execution within a processor is very quick and inexpensive for the time. Task parallelism is shown in figure 3.

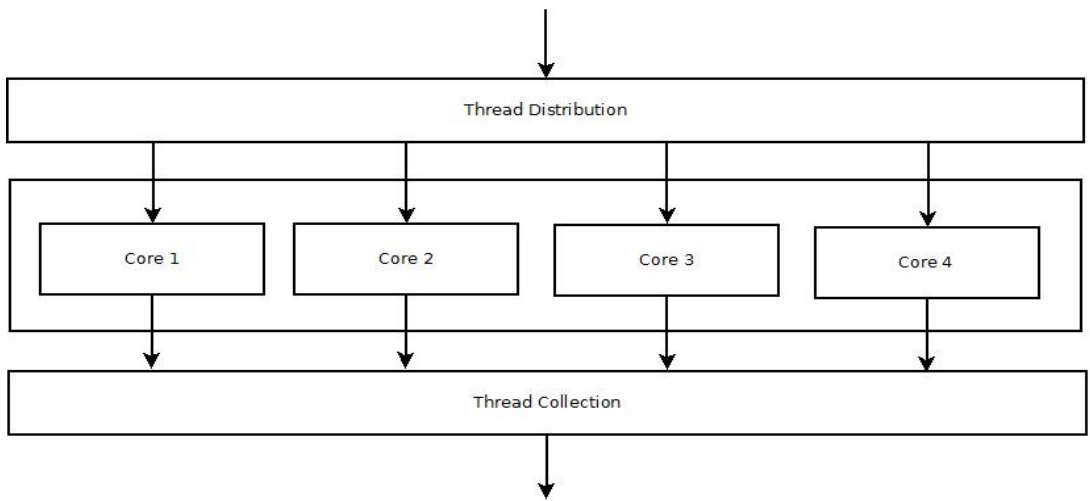

Figure 3. Multithreading in multicore systems

The microprocessor has only CPU, it does not have RAM, ROM and other peripheral on the chip, but the microcontroller contain all the basic components as shown in figure 4. A microcontroller has two kinds of design mechanics, i.e. UMA (Unified Memory Architecture) and NUMA (Non Unified Memory Architecture), although it can also be distinguished by Single Instruction Multiple Data (SIMD) and Multiple Instruction Multiple Data(MIMD). Table 2 illustrated the difference between microcontroller and microprocessor which show that the microcontroller has more advantage over microprocessor.

Table 2. Difference between microcontroller and microprocessor

| Parameter | Microcontroller | Microprocessor |
|---|---|---|
| Meaning | Computer on chip | CPU on chip |
| Inbuilt components | CPU, Memory & Peripheral Unit | CPU ONLY |
| Memory Architecture | UMA | UMA and NUMA |
| Purpose | Specific | General |
| Chips | All components on chip | Multiple chip for components |
| Circuit | Simple | Complex |
| No of Registers | More | Less |
| Operation | Registers based | Memory based |
| Clock Frequency | Low | High |
| Cost | Low | High |
| Speed | Fast | Fastest |

.
There are vast applications of multicore microprocessors. The market trends show that in general purpose computing processors are playing a key role. The entire major manufacturer is deploying their Integrated Circuit(IC) with its functionality. Processors are designed for general purpose; it can't be used in embedded application for that user need multicore microcontroller. Multicore microcontroller [20] is used in embedded application and Real Time Operating System(RTOS) in which time is a major factor. In real time application multicore microcontroller play a major role. In all major time critical operations, including defence, military, medical, industrial, etc. handled by multicore microcontroller. Recently mangalyaan launched in Indiain Mars orbit in low cost is managed by multicore microcontroller. Table 3 compares the multicore microcontroller and multicore microprocessor and show that multicore microcontroller has a huge advantage over multicore microprocessor.

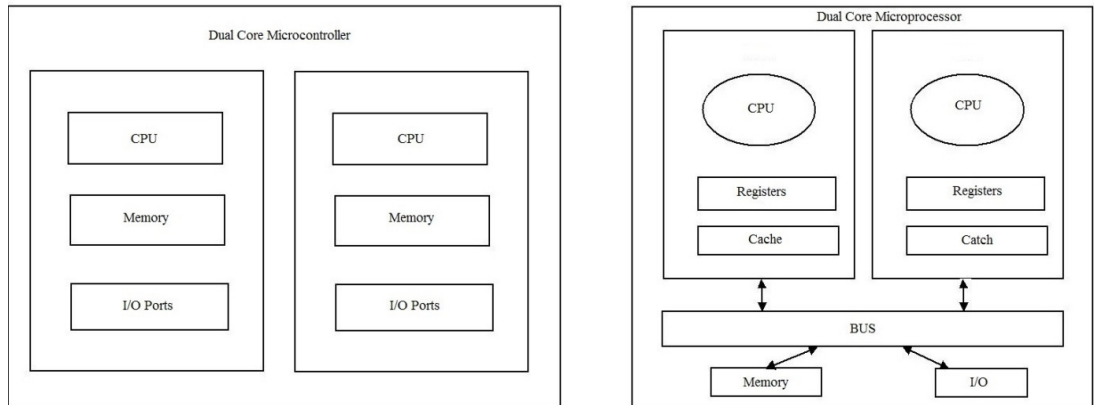

Figure 4. Illustrating Multicore Microcontroller vs Multicore Microprocessor

Table 3. Distinction between multicore microcontroller and multicore microprocessor

| Parameters | Multicore Microcontroller | Multicore Microprocessor |
|---|---|---|
| Architecture | Harvard | Von Neumann |
| Instruction Set | RISC | CISC |
| Power | M Hz | G Hz |
| Execution time | In nano sec or micro sec | In milisec or in sec |
| Cost | Chip | Expensive |
| Interrupt | Given by the program | Hardware and Software Interrupt |
| Priority | Not define, All are executed parallel | Masskable and Non Maskable |
| Cache | Not used | Used |
| Tile | Define | Not define |
| Time Critical Analysis | Can be done | Can not done |
| Power conception | Low | High |
| Power Saving Mode | Available | Not Available |
| Application | Embedded system and RTOS | General purpose |

# 2. XMOS® startKIT

startKIT [21] is a xCORE multicore microcontroller that has eight 32 bit logical processor cores on two tiles as shown in figure5 which is taken from XMOS® XS architecture [22]. The size of startKIT is very small. The startKIT dimensions are 94 x 50mm. The startKIT require 5V which are given by Micro-USB cable. The regulator is used to convert this 5V to 1V and 3V which is used by external devices. Table 4 gives the overview of startKIT. It is very easy to use and simple to program. User can easily design complex embedded system using high level language. Each core acts as separately and able to run multiple real time tasks simultaneously. It provides 500 million instructions per second (MIPS) which make this more powerful than conventional microcontroller. It provides a uniquely scalable, timing deterministic architecture that provides extremely low latency and an I/O response that is 100 times faster than standard processors. If a core is waiting for data, the xTIME hardware scheduler will pass the execution resource to the next core making efficient use of the available processor resource and saving power.

XMOS® provides xTIMEcomposer [23] IDE for designing applications. It supports high level languages like C,C++, XC (extension of C). It has inbuilt debugger, compiler, simulator and editor. xTIMEcomposer provide many functionalities to improve the performance and check various parameters like time, delay. xTIMEcomposer provide XMOS® Timing Analyser (XTA [24]) tool which allow developers to identify worst case and best case execution time for code blocks and functions. It also provides xSCOPE [25] which allow capturing the data from running time. Figure 6 explains the architecture of XMOS® startKIT. startKIT has two tiles. Tile 0 is dedicated to the integrated debugger and Tile 1 is user-programmable. It has 8 cores with 500 MIPS. Micro-USB connector (B) is used as a debugger. It connects to the host PC and allows running the program.

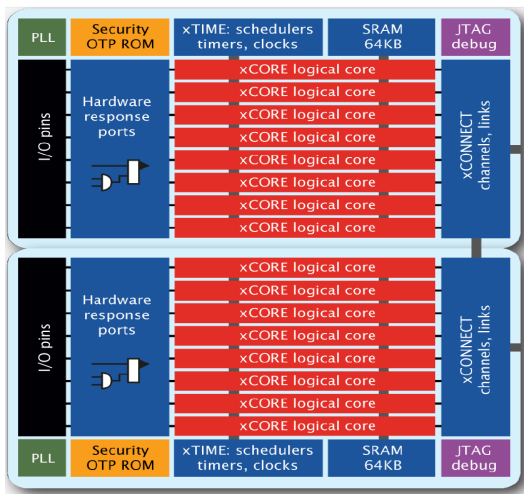

| Dimension | 94 x 50mm |
|---|---|
| No of Core | 8 |
| No of Tiles | 2 |
| Word Length | 32 Bit |
| Architecture | RISC |
| Cache | Not Used |
| Clock Freq. | 500 MHz |
| SRAM | 64 KB |
| FLASH | 256K Bytes |
| Voltage | 5V, 3V3, GND |
| Programming Language | High Level Language |

Table 4. XMOS® startKIT overview                Figure 5. XMOS® architecture [22]

startKIT provide GPIO pins which allow developers to reconfigure the capabilities of devices to support many different applications. PCIe slot is used to extend the hardware capabilities, 1*5 PCIs connected (J6) and 1*12 GPIO header (J7) is used to connect external hardware or peripheral. If PCIs slot is not used, then the user can use GPIO header [26]. The startKIT is compatible with the Raspberry Pi connection. The developer can connect Raspberry Pi board with startKIT using 2*13 Raspberry Pi headers. It is compatible with Raspberry Pi [26] connection. It disables the LED and push button so the usercannot use LED and button in the Raspberry Pi header. XMOS® Links (E)1x13 pin GPIO header (J8) used for connecting multiple startKITs together.The startKIT provides two and four-zone capacitive touch slider which is illustrated as an F in figure 6.The startKIT has nine 3*3 green LED as shown in section G.ThestartKIT has two additional green LED as a display in H.It has 256 KBytesof Serial Peripheral Interface (SPI) FLASH memory (I), which can be configured by the program.startKIT has one push button (J) which used as input and its status can be checked by software.startKIT provide 2*3 analogy input header (K) which is used to give analogyinput.Tile 1 is clocked at 500 MHz, and the I/O ports are 100MHz. The startKIT board is clocked at 24MHz by a crystal oscillator as shown in L.

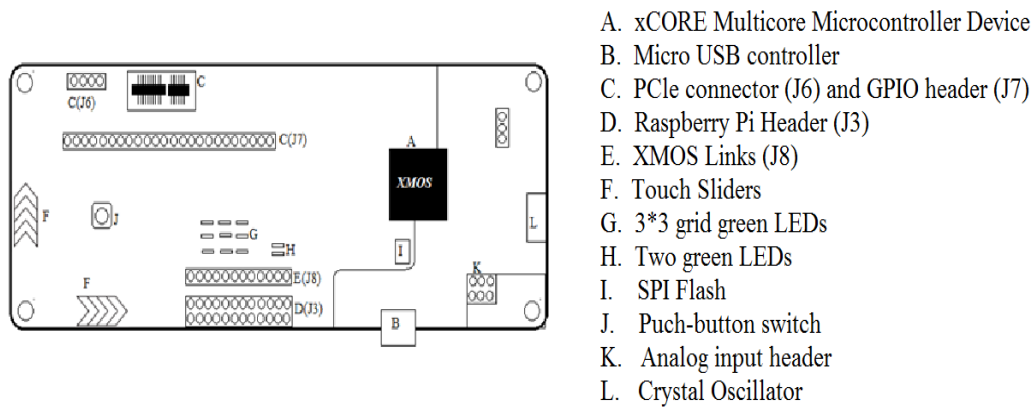

A. xCORE Multicore Microcontroller Device
B. Micro USB controller
C. PCIe connector (J6) and GPIO header (J7)
D. Raspberry Pi Header (J3)
E. XMOS Links (J8)
F. Touch Sliders
G. 3*3 grid green LEDs
H. Two green LEDs
I. SPI Flash
J. Puch-button switch
K. Analog input header
L. Crystal Oscillator

Figure 6. Architecture [24] of XMOS® startKIT

## 2.1. TASK PARALLELIZATION IN XMOS® STARTKITARCHITECTURE

XMOS® [27] use 'par'constructs which allow several tasks to run parallel as illustrated in figure 7 (a) [27]. The compiler automatically checks how many cores are used and allocatesthe one task to one core whichever is free. The user can assign the task based on event occurs. XMOS® used 'select' keyword as illustrated in figure 7 (b) [27] which pause the task and wait for the event to occur and handle the event which isoccurring which is used as an interrupt. The event is assigned by the function of interrupt service routine (ISR). Multicore system also allows channelling and time stamping.Time stamping manages multiple events with a microcontroller that all require different timing. For example, you might want to control a servomotor (which requires a 20 millisecond delay), blink an LED once a second, and read some sensors (which should be read as frequently as possible. One way to handle this is to keep track of a time stamp for each event. The channel is used to communicate between the tasks. The user performs tasks parallel with shared memory. It creates a hang up situations when both the process tries to change or use the same data at the same time. Channel [27] resolves this issue. It sends the signal to processor that presenttask is finished now and so anothertask can be performed. XMOS® used 'chan' keyword as shown in figure 7 (c) [27]. Sometimes microcontroller can be ina busy state,at that time user has to wait for some time. XMOS®time stamping explain in figure 7 (d) [27] in which after the 100 time units led will be high and after the wait for 100 time unit led will on low stage.

```
int main (void)
{
        par
        {
                task1();
                task2();
                task3();
                task4();
        }
}
```

Figure 7(a)

```
select {
case event1 :
// handle the event
...
break ;
case event2 :
// handle the event
...
break ;
}
```

Figure 7(b)

```
chan c;
par
{
        task1 (c);
        task2 (c);
}
```

Figure 7(c)

```
led @ 100 <: 1
led @ 100 <: 0
```

FIgure 7(d)

Figure 7. Time critical analysis and multitasking using XMOS® startKIT

## 3. EXPERIMENTAL RESULTS

Experimental results mentioned in this section is based on time critical analysis of sequential and parallel task on multi core microcontroller XMOS® and the multicore microprocessor i3 and i5.

### 3.1. Time critical analysis of sequential and parallel tasks on multicore microprocessor

Analysis of the sequential and parallel [28] behaviors are observed that there is a huge difference in their execution time. Multithread program [29] in c executed using POSIX and the thread level parallelism (TLP) is shown in figure 8. First, it is triedwith two threads then four and eight threads are usedand shown in figure 8 which is just giving the status of the tread. The results are checked with the help of profiling, which is obtainedatthe table5. The same process is applied to a different machine.

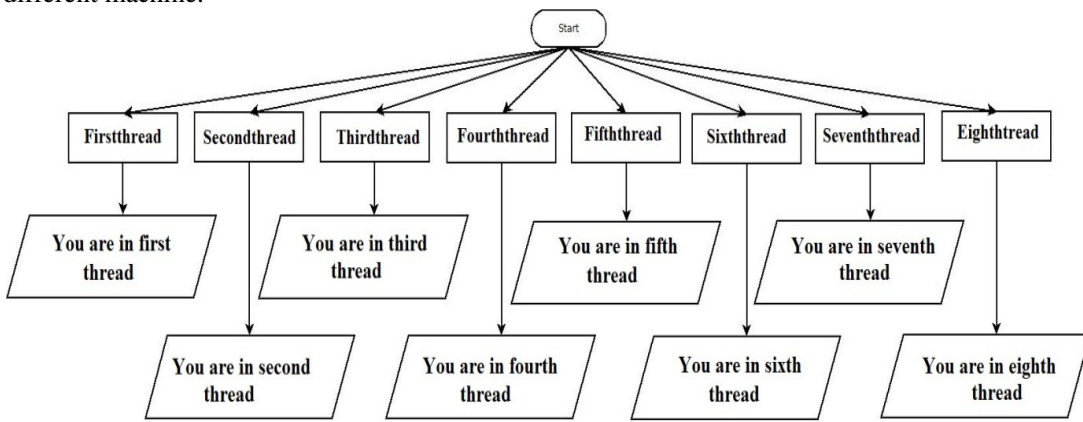

Figure 8. Flow chart of thread level parallelism

Nothe same thingng is done by simple sequential c code as shown in figure 9 and obtain the sequential table as illustrated in table 6 afteanalyzingng the each function execution time as well aprogramme execution time. The same analysisis done with various otherarchitectures [30] andthe difference in their execution time and the cache misses by the processor are examined First program load in the cache then it will be executed. If any variable's value or any instruction is not available in cache [31] then processor first fetches the instruction which takes some time that is called cache misses.

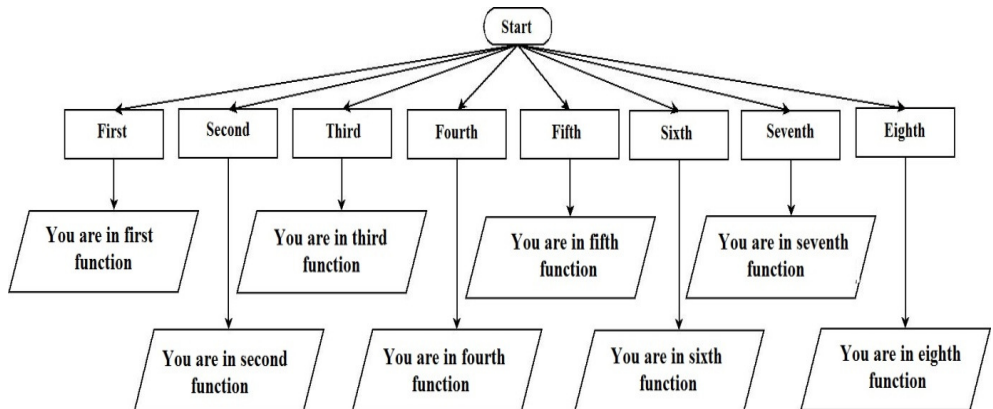

Figure9. Flow chart of sequential program

Table 5 Multi-thread Profiling

| Architecture | Compiler | No of Threads | Compile time(sec) | Exc. time of each function(sec) | Cache Misses (%) |
|---|---|---|---|---|---|
| Intel i3 | Gcc | 2 | real =5.002<br>user =9.976<br>sys =0.022 | Firstthread = 5.64<br>Secondthread =4.44 | I1 Misses =0.48<br>D1 Misses =1.1<br>LLi Misses =0.29<br>LLd Misses =0.6<br>LL Misses =0.3 |
| Intel i5 | Gcc | 2 | real =5.002<br>user =5.580<br>sys =0.003 | Firstthread = 1.18<br>Secondthread =0.21 | I1 Misses =0.46<br>D1 Misses =1.1<br>LLi Misses =0.25<br>LLd Misses =0.6<br>LL Misses =0.3 |
| Intel i3 | Gcc | 4 | real =5.002<br>user =18.481<br>sys =0.0721 | Firstthread =4.57<br>Secondthread =3.54<br>Thirdthread =2.42<br>Fourththread =2.04 | I1 Misses =0.49<br>D1 Misses =1.1<br>LLi Misses =0.25<br>LLd Misses =0.6<br>LL Misses =0.3 |
| Intel i5 | Gcc | 4 | real =5.002<br>user =16.712<br>sys =0.004 | Firstthread =1.28<br>Secondthread =1.21<br>Thirdthread =1.06<br>Fourththread =0.95 | I1 Misses =0.48<br>D1 Misses =1.1<br>LLi Misses =0.20<br>LLd Misses =0.6<br>LL Misses =0.3 |
| Intel i3 | Gcc | 8 | real =5.003<br>user =21.538<br>sys =0.052 | Firstthread =1.25<br>Secondthread =1.11<br>Thirdthread =1.03<br>Fourththread =0.79<br>Fifththread=0.67<br>Sixththread=0.50<br>Sevenththread=0.22<br>Eighththread=0.11 | I1 Misses =0.50<br>D1 Misses =1.1<br>LLi Misses =0.25<br>LLd Misses =0.6<br>LL Misses =0.3 |
| Intel i5 | Gcc | 8 | real =5.003<br>user =19.948<br>sys =0.004 | Firstthread = 2.70<br>Secondthread =1.29<br>Thirdthread =0.19<br>Fourththread =0.64<br>Fifththread=0.64<br>Sixththread=0.73<br>Sevenththread=0.65<br>Eighththread=0.18 | I1 Misses =0.49<br>D1 Misses =1.1<br>LLi Misses =0.24<br>LLd Misses =0.6<br>LL Misses =0.3 |

Table6. Sequential Profiling Table

| Architecture | Compiler | No of Functions | Compile time(sec) | Exc. time of each function(sec) | Cache Misses (%) |
|---|---|---|---|---|---|
| Intel i3 | Gcc | 2 | real =30.528<br>user =30.480<br>sys =0.044 | First = 15.40<br>Second =15.41 | I1 Misses =0.49<br>D1 Misses =1.1<br>LLi Misses =0.25<br>LLd Misses =0.6<br>LL Misses =0.3 |
| Intel i5 | Gcc | 2 | real =18.579<br>user =18.580<br>sys =0.008 | First = 9.14<br>Second =9.13 | I1 Misses =0.48<br>D1 Misses =1.1<br>LLi Misses =0.25<br>LLd Misses =0.6<br>LL Misses =0.3 |
| Intel i3 | Gcc | 4 | real =1m0.987<br>user =1m0.911<br>sys =0m0.068 | First = 15.39<br>Second =15.41<br>Third =15.40<br>Fourth =15.30 | I1 Misses =0.48<br>D1 Misses =1.1<br>LLi Misses =0.25<br>LLd Misses =0.6<br>LL Misses =0.3 |
| Intel i5 | Gcc | 4 | real =37.332<br>user =37.340<br>sys =0.016 | First = 9.47<br>Second =9.28<br>Third =9.22<br>Fourth =9.22 | I1 Misses =0.46<br>D1 Misses =1.1<br>LLi Misses =0.25<br>LLd Misses =0.6<br>LL Misses =0.3 |
| Intel i3 | Gcc | 8 | real =2m2.215<br>user =2m2.030<br>sys =0m0.160 | First = 15.39<br>Second =15.34<br>Third =15.32<br>Fourth =15.31<br>Fifth=15.34<br>Sixth=15.32<br>Seventh=15.34<br>Eighth=15.32 | I1 Misses =0.48<br>D1 Misses =1.1<br>LLi Misses =0.25<br>LLd Misses =0.6<br>LL Misses =0.3 |
| Intel i5 | Gcc | 8 | real =1m13.649<br>user =1m13.692<br>sys =0.024 | First = 9.27<br>Second =9.27<br>Third =9.26<br>Fourth =9.26<br>Fifth=9.25<br>Sixth=9.24<br>Seventh=9.22<br>Eighth=9.13 | I1 Misses =0.48<br>D1 Misses =1.1<br>LLi Misses =0.25<br>LLd Misses =0.6<br>LL Misses =0.3 |

Various architectures used different frequency which affects the execution time. The experimental results mentioned in table 5 and table 6 explain how function execution time varies with architecture which is illustrated in figure 10. Not only the architecture but also the number of threads [32] is important the execution time of a machine. In parallelcomputation all the CPUs are treated as individual entity which are connected to each other for better communication and it executes all the threads at the same time, which could be proven with the help of figure 11, it could also be observed with the same diagram that every function has a different time boundation that is also known as time bounded computation.The execution speed of Intel® i3 to Intel® [33] i5 reduces as the frequency of the device increases its results are illustrated in table 6.Aftercomparing the results mentioned in tables 5 and 9 the conclusion is that the good amount of possibility that multicore arc. For controllers are very effective and efficient for time critical execution.

Figure 10 shows the execution time depends upon the architecture also. As Intel[®] i5 has higher frequency [34] than Intel[®] i3 and also Intel[®] i3 has only 2 cores (dual core) and Intel[®] i5 has four courses. Intel[®] i5 has less execution time in comparison to Intel[®] i3. Execution time reduces as the frequency and number of core increases. Even applying the same code for different-2 processors, then also we get different-2 time. As it is noticed that in every caseexecution time is less in parallel of Intel[®] i5 than Intel[®] i3 (which has number of cores) as shown in figure 11 that's proves  requirement of multi-core and parallel processing is increasing rapidly.

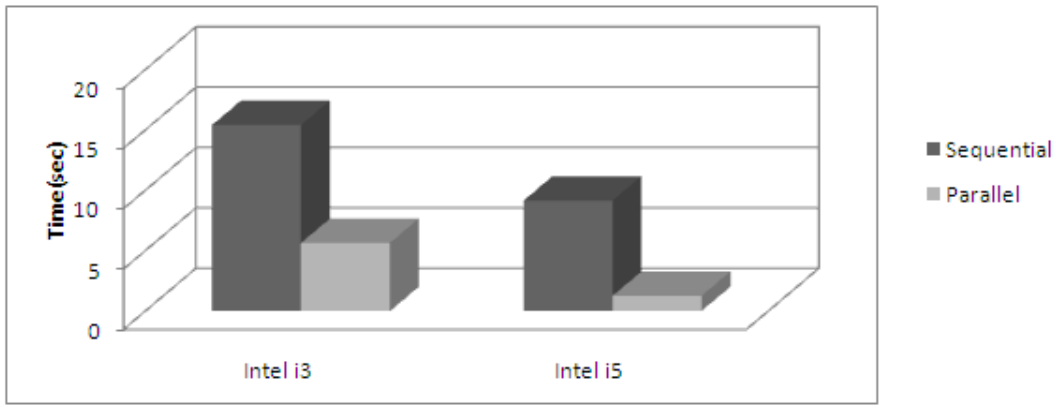

Figure10. Function Execution time

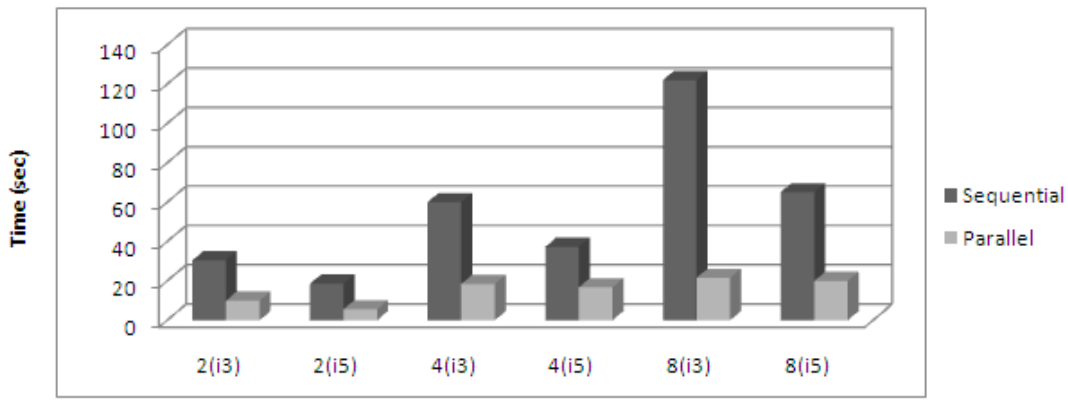

Figure 11. Total execution time

Although same program is used for sequential as well as the parallel processing, but in the parallel processing Intel[®] i3 takes almost 5 sec and Intel[®] i5 takes 2 sec and in case of sequential program Intel[®] i3 takes 15 sec and Intel[®] i5 takes 9 sec which is shown in figure10. The total execution time [35] depends upon the architecture and the number of threads or function used by the program. As the number of threads and function increases, total execution time of the processor increases as we used same threads and the same program. Using [®]i3 with 2 functions in a parallel program is given worst execution time, whereas in Intel[®] i5 with same 2 functions in parallel processing [36] gives best result as illustrated in figure 11.

## 3.2. Time Critical analysis of sequential and parallel task on Multi core microcontroller

Results show that the development of parallel program by xtime composer for SK131552, the task mentioned in the core has the special property of activating and setting a specific character using 3x3 led matrix which is communicating through inbuilt com port i.e. 32 which is 20 bits long. The 1st bit from first led & 2nd bit from second led & 3rd bit assign to third led. 8th bit, 9th, 10th, 11th, 12th, 13th bit assigns to fourth, fifth, sixth, seventh, eighth & ninth led respectively, Others are don't care. The words like 'Y' 'O' 'U' 'K' is as shown in figure 12(a)-(d).

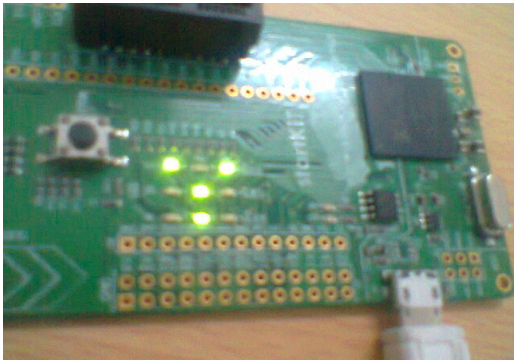

Figure 12(a). Display Y 3*3 LED

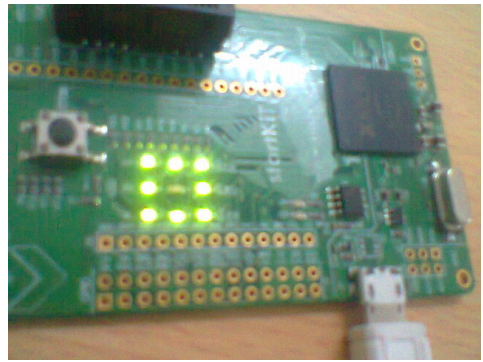

Figure 12(b). Display O 3*3 LED

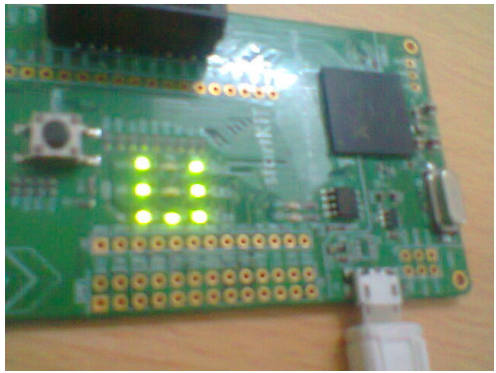

Figure 12(c). Display U 3*3 LED

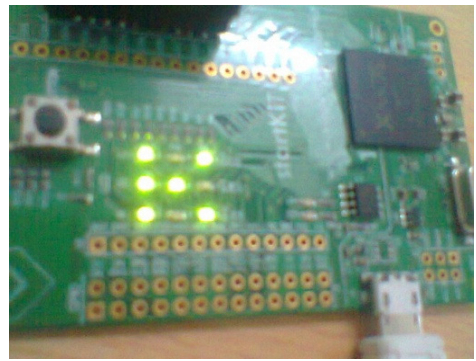

Figure 12(d). Display K 3*3 LED

The sequential flow chart of spinning ball is shown in Figure 13 in which 3x3 led glows and rotate in a circle.Figure 14 shows the flow chart of a pattern shift in which LEDs of the first column, then second, then third column, then first, second then third row glows and are repeated again in the same manner.

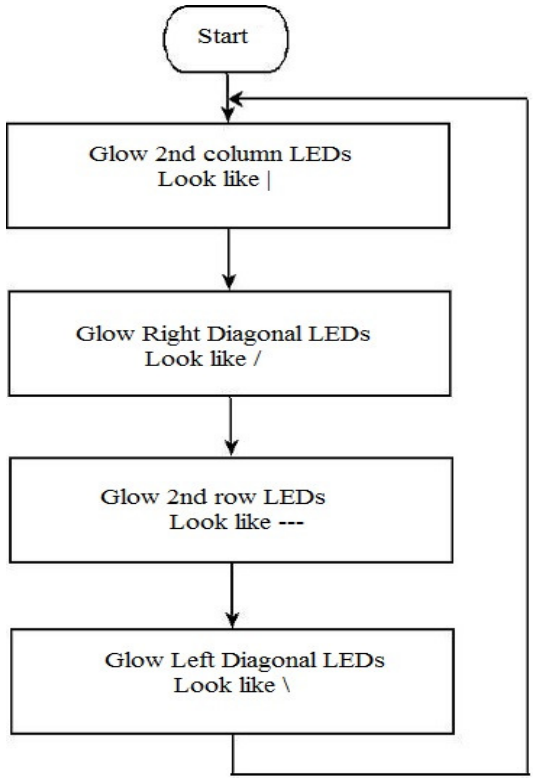

Figure 13. Sequential execution of spinning ball program

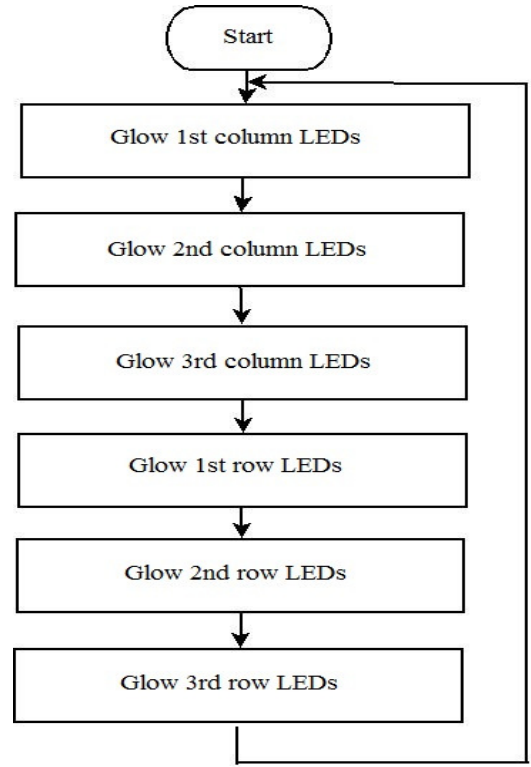

Figure 14. Sequential execution of pattern formation program

Parallelization is demonstrated using the servo motor. Flow chart of Servo motor control is shown in figure 15 using xTOOLS [27] the obtained result is shown in figure 16. The platform summary shows the how many cores, timer, memory and channel used by the program. It also displays the value of timer and how much memory occupied by the program. startKIT [27] has 8 logical cores, 32 channels, 655036 bites memory space and 10 timer. Table 7 illustrates how many of them are occupied by this program.

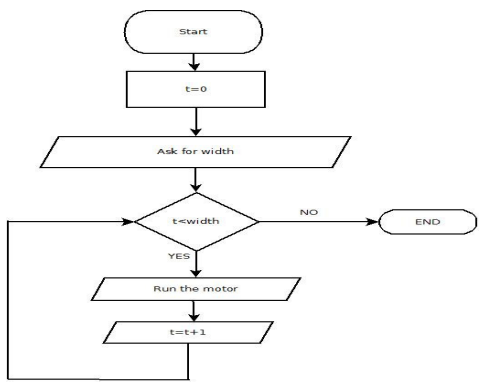

Figure 15. Flow chart of servo motor control

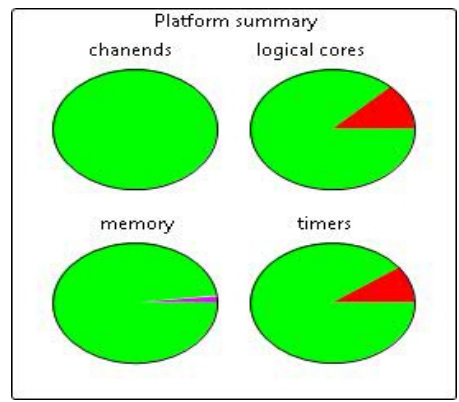

Figure16. Analysis

Table 7. Result of servo motor control

| Name | Used | Free |
|------|------|------|
| Chanends | 0% | 100% |
| Logical Cores | 1(12.5%) | 7(97.5%) |
| Memory | 336bit(0.51%) | 65200bit(99.49%) |
| Timer | 1(10%) | 9(90%) |

Multitasking [36] is also performed in this which 4 tasks perform simultaneously as shown in figure 17, with the help of 'par' keyword. First two cores are assigned to blink the two additional led and third core is assigned to display '+' and fourth core is used to display 'X' in 3x3 led. Third and fourth core work on same 3x3 led so it required channel by using 'chan' keyword to assign the task so that both cannot overlap also profiling is done using gproof and noted that how much time is taken by all the functions and result [37] is shown in table 8 and its profiling graph is illustrated in figure 18. How much memory, channels, logical cores and timer used by this program is illustrated in table 9. In this program stack is also used so memory is occupied by stacking as well as program.

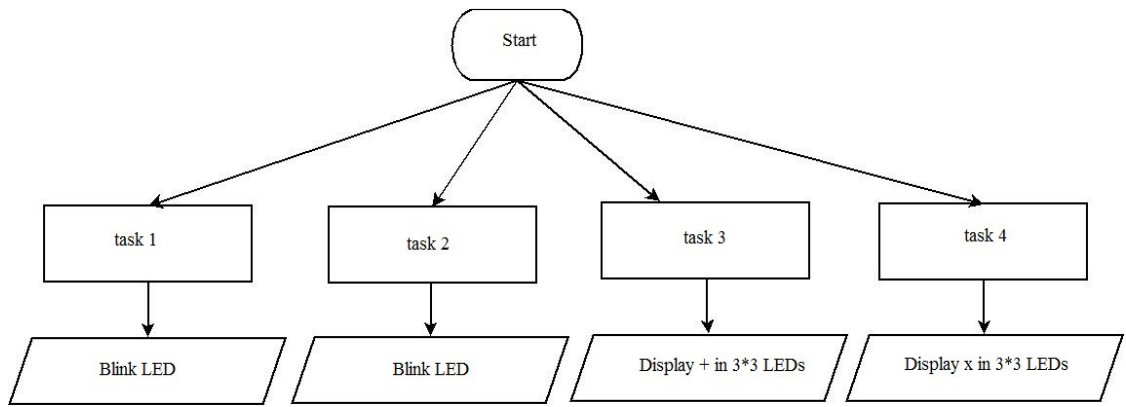

Figure17. Flow chart of multitasking

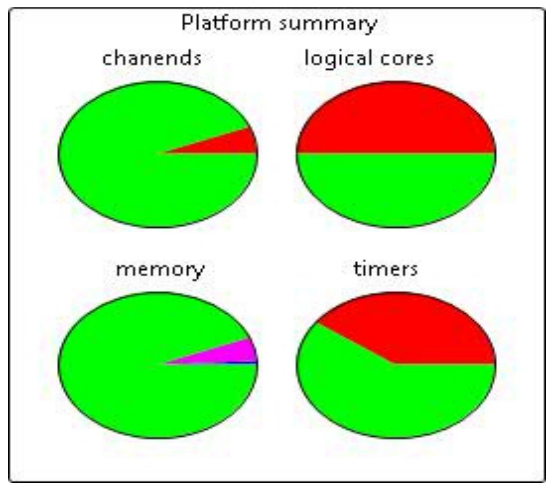

Table 8. Execution time of each thread

| Function Name | Time |
|---|---|
| task1 | 125ns |
| task2 | 152ns |
| task3 | 2.548 micro sec |
| task4 | 3.45 micro sec |

Table 9. Result of multitasking

| Name | Used | Free |
|---|---|---|
| Chanends | 3(9.38%) | 29(90.63%) |
| Logical Cores | 4(50%) | 4(50%) |
| Memory(Stack) | 604(0.92%) | 60812(92.79%) |
| Memory(Program) | 4120(6.29%) | |
| Timers | 4(40%) | 6(60%) |

From the experimental results we can illustrate that the time taken by the sequential programming on multicore[38] microcontroller is in 'micro seconds[39] whereas time [39] taken by the multicore microprocessor is in 'seconds' and 'minutes' which proves that the sequential task on microcontroller can be done faster than the microprocessor. Similarly, the time taken for task parallelization on multicore [40] microcontroller is in 'nanoseconds' whereas multicore microprocessor takes 'seconds' which again proves that the parallel task on microcontroller can be done faster than the microprocessor.

# 4. CONCLUSIONS

It can be seen that  the time critical analysis on micro controller, task parallelization for complex programming structures can be compiled in less time, although it has been observed that these techniques of programming are easy for an advanced developer but  tough for a new or the beginners. Using above mentioned results user easily analyze the performance of multicore microcontroller. The task Parallelization for multicore controller is still a challenge due to its complex programming structure, but it gives high versatility and reliability to perform real-time operations. The Future focus of the research is to implement it in automation control system using SPI and I2C interfaces.In this research every thread is assigned by the controller to its different cores by itself, future goal will concentrate by hand on assignment of threads on the cores. Also in this paper time critical analysis results and examples are restricted only for basic applications.Future work will be focused on expansion of present experimentations on time critical analysis on real time applications and embedded system.

# ACKNOWLEDGEMENTS

The authors would like to thank to Dr. Abhishek Sharma for their guidance and constant supervision as well as for providing necessary information regarding the project & also for their support in completing the project. The authors would like to express gratitude towards parents &

member of LNMIIT for their kind co-operation and encouragement which help in completion of this project. Thanks and appreciations also go to the Texas instrumental Lab.

## AUTHORS

**Prerna Saini** is pursuing B. Tech in the branch Electronics and Communication Engineering at the LNM Institute of Information Technology, Jaipur (Rajasthan), India. Her area of interests is embedded System, Multicore Microcontroller and Microprocessor and High Performance Computing.

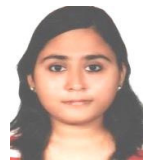

**Ankit Bansal** is pursuing B. Tech in the branch Electronics and Communication Engineering at GLA University, Mathura (U.P), India. His current interests are in Embedded System, Multicore Microcontroller and Microprocessor and cloud computing.

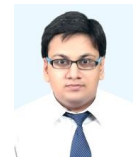

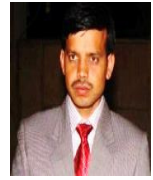

**Abhishek Sharma** obtained his B. Tech degree in ECE from Jiwaji University, Gwalior, India. He worked with two Telecom Companies i.e. ZTE and Huawei. After he worked in CNR, Milan as Research Engineer. He finished his Ph. D in multicore and many core systems in 2010 from university of Genoa, Italy. He is currently working  as Assistant Professor in the Dept. of ECE in LNMIIT, Jaipur (Rajasthan), India. His current research interest includes high performance embedded system, many cores and multicore architecture.